\documentclass[10pt,conference]{IEEEtran}
\IEEEoverridecommandlockouts
\usepackage[utf8]{inputenc}
\usepackage{bm}
\usepackage{amsfonts}
\usepackage{amssymb}
\usepackage{amsmath}
\usepackage{algorithm}
\usepackage{algpseudocode}
\usepackage{graphicx}
\usepackage{siunitx}
\usepackage{cite}
\usepackage[font=small]{caption}
\usepackage{multicol}

\usepackage{lipsum}
\usepackage[hidelinks]{hyperref}

\usepackage[letterpaper,top=0.7in,right=0.65in,left=0.65in,bottom=1.05in]{geometry}
\begin{document}

% Copyright
\onecolumn
\thispagestyle{empty}
\Huge IEEE Copyright Notice \\

\large
© 2025 IEEE. Personal use of this material is permitted. Permission from IEEE must be obtained for all other uses, in any current or future media, including reprinting/republishing this material for advertising or promotional purposes, creating new collective works, for resale or redistribution to servers or lists, or reuse of any copyrighted component of this work in other works.

\vfill
This work has been accepted for publication in \textit{2025 IEEE Wireless Communications and Networking Conference (WCNC)}. The final published version is available via IEEE Xplore, DOI: 10.1109/WCNC61545.2025.10978323.

\clearpage
\twocolumn
\normalsize

% Article
\title{Coexistence of eMBB+ and mMTC+ in \\ Uplink Cell-Free Massive MIMO Networks
\thanks{This paper was carried out within the frameworks of the Horizon Europe project CENTRIC (Grant No. 101096379) and the Horizon Europe Programme Marie Skłodowska-Curie Actions (MSCA) Postdoctoral Fellowships (PF) - European Fellowships (EF) DIRACFEC (Grant No. 101108043). The European Union also supported the work of Stefano Buzzi under the Italian National Recovery and Resilience Plan (NRRP) of NextGenerationEU, partnership on ``Telecommunications of the Future'' (PE00000001 - program ``RESTART''), structural project 6GWINET. Views and opinions expressed are those of the author(s) only and do not necessarily reflect those of the European Union. The European Union cannot be held responsible for them.}}

\author{\IEEEauthorblockN{Sergi Liesegang$^{1,2}$ and Stefano Buzzi$^{1,2,3}$}
\IEEEauthorblockA{$^1$\textit{DIEI, Università degli Studi di Cassino e del Lazio Meridionale, 03043 Cassino (FR) -- Italia} \\
$^2$\textit{Consorzio Nazionale Interuniversitario per le Telecomunicazioni, 43124 Parma (PR) -- Italia} \\ 
$^3$\textit{DEIB, Politecnico di Milano, 20122 Milano (MI) -- Italia}}
E-mails: \{sergi.liesegang, buzzi\}@unicas.it}

\maketitle

\begin{abstract}
This paper tackles the problem of designing proper uplink multiple access schemes for coexistence between enhanced mobile broadband+ (eMBB+) users and massive machine-type communications+ (mMTC+) devices in a terminal-centric cell-free massive MIMO system. Specifically, the use of a time-frequency spreading technique for the mMTC+ devices has been proposed. Coupled with the assumption of imperfect channel knowledge, closed-form bounds of the achievable (ergodic) rate for the two data services are derived. Using suitable power control mechanisms, we show it is possible to efficiently multiplex eMBB+ and mMTC+ traffic in the same time-frequency resource grid. Numerical experiments reveal interesting trade-offs in the selection of the spreading gain and the number of serving access points within the system. Results also demonstrate that the performance of the mMTC+ devices is slightly affected by the presence of the eMBB+ users. Overall, our approach can endow good quality of service to both 6G cornerstones at once.
\end{abstract}


\section{Introduction} \label{sec:1}
During this decade, academy and industry are devoted to the evolution of the sixth generation of cellular networks (6G) \cite{Hua22}. The exponential growth in the number of terminals and the incessant demand for heterogeneous data services, evinced the urge to develop novel solutions \cite{Eri23}. 6G will extend its predecessor use cases: enhanced mobile broadband+ (eMBB+), which pursues high rates; ultra-reliable low latency communications+, which seek short delays and small error probabilities; and massive machine-type communications+ (mMTC+), which need vast connectivity and low power consumptions. This difference in requirements makes providing simultaneous support to all cornerstones a challenging problem.

\textit{Coexistence} will then be a pivotal factor in the design and implementation of future mobile systems. New technologies like millimeter-wave bands; centralized and distributed large-scale MIMO (multiple-input multiple-output); reconfigurable intelligent surfaces; and edge intelligence emerge as potential candidates to overcome the aforementioned issues \cite{Ngu22}. 

In past releases, equipping (macro) base stations with lots of antennas, i.e., massive MIMO (mMIMO) \cite{Mar10}, has permitted operators to increase the rates. However, the required circuitry (and consumption) scales with the number of antennas, which quickly becomes prohibitive. Such centralized approaches also fail to ensure good quality of service (QoS) to far away users (due to poor channel conditions and inter-cell interference). 

To tackle such problems, a promising solution is to deploy the antennas, in the form of access points (APs), across the scenario and design the network in a \textit{user-centric} manner. This translates into a distributed architecture without cell borders, known as \textit{cell-free} mMIMO (CF-mMIMO), that has been shown to enhance the coverage (or QoS) and reduce the power consumption of its collocated counterparts \cite{Int19}.

Multi-antenna technology can also pave the way to heterogeneous networks (i.e., interference among data services can be mitigated with the resulting spatial diversity). However, the coexistence of human and machine communications is (often) addressed via orthogonal multiple access (MA), which might be outperformed by non-orthogonal solutions \cite{Yan19}.

With the above considerations, our purpose is to fill this ``coexistence" gap: conceive suitable MA schemes that reduce the impact of the interference between services (not available to date). This work focuses on an uplink (UL) CF setup where eMBB+ users and mMTC+ devices transmit employing shared resources. Since mMTC+ are characterized by low data rates, it is here proposed to underlay them as spread-spectrum signals \cite{Sim94} in the time-frequency grid used by the eMBB+ users. This MA method will help to significantly mitigate the interference among services thanks to the large duration and low transmit power of the resulting mMTC+ packets \cite{Mol22}.

The lack of channel knowledge is also incorporated in our study: we derive closed-form lower bounds for the data rates with statistical information only. Note that, when decorrelating the mMTC+ signature sequences, several (cumbersome) $8$-th order moments appear, for which analytic expressions are also obtained. As for performance evaluation, we will later design the power allocation under a fair policy: maximize the eMBB+ rates subject to QoS constraints on the mMTC+. To the best of our knowledge, no similar contributions have been reported.

The remainder of this paper is organized as follows. Section~\ref{sec:2} introduces the system model. Section~\ref{sec:3} is devoted to the achievable rates. Section~\ref{sec:4} formulates and solves the QoS optimization problem. Section~\ref{sec:5} validates the results through numerical experiments. Section~\ref{sec:6} concludes the work.

\section{System Model} \label{sec:2}
\begin{figure}[t]
\centerline{\includegraphics[scale=0.245]{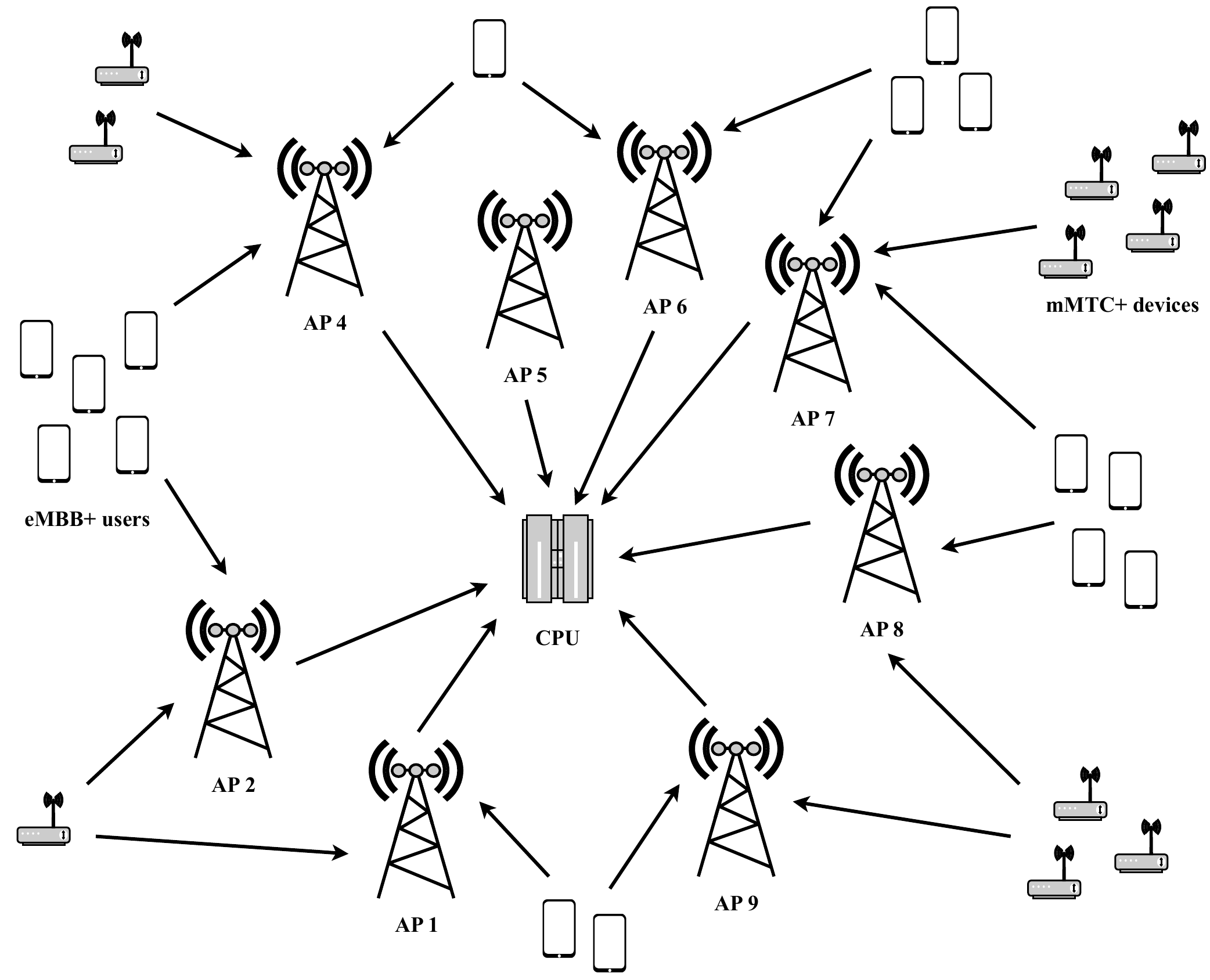}}
\caption{Illustrative example of a terminal-centric CF-mMIMO deployment, where $M = 9$ APs with $L = 3$ antennas serve $K_u = 15$ eMBB+ users and $K_d = 10$ mMTC+ devices simultaneously.}
\label{fig:1}
\vspace{-5mm}
\end{figure}

Throughout this work, we explore a CF scenario akin to that described in \cite{Lie24}, where $M$ APs equipped with $L$ antennas are connected via a high-capacity fronthaul to a central processing unit (CPU) and serve $K_u$ single-antenna eMBB+ users. Unlike \cite{Lie24}, now $K_d$ single-antenna mMTC+ devices will coexist with the mobile users. This configuration is illustrated in Fig.~\ref{fig:1}, indicating that the different terminals\footnote{For the sake of brevity, in this paper, the concept of ``terminals'' might indistinguishably refer to both eMBB+ users and mMTC+ devices.} transmit to only a subset of APs \cite{Elw23}, i.e., the notion of user-centric architectures is thus extended to \textit{terminal-centric} deployments.

Recall that users and devices\footnote{To further maintain conciseness, the terms ``users'' and ``devices'' will be used to solely designate eMBB+ and mMTC+ terminals, respectively.} share all the available resources. More precisely, in line with 3GPP terminology \cite{3GPP36814}, we consider a grid comprising $N$ physical resource blocks (PRB) of time-frequency samples. To facilitate the coexistence of both services, we propose a spread-spectrum MA for the mMTC+ \cite{Sim94}. This allows us to separate their messages easily and, as discussed later, also mitigate interference towards the rest of mobile users. Additionally, it is worth noting that, as opposed to eMBB+, where high rates are expected, mMTC+ devices are usually battery-constrained and require high energy efficiencies. Hence, the resulting \textit{spreading gain} can contribute to reducing peaks of power consumption \cite{Mol22}. 

All this is presented in the following, where we elaborate more on the two stages of UL communication: channel estimation and data transmission. Before that, we will dedicate a subsection to discuss the propagation model.

\subsection{Propagation Channel}\label{sec:2.1}
Assuming channel stationarity over the resource grid, the link from user $u$ to AP $m$ at PRB (or epoch) $n$ is \cite{Int23}
\begin{equation}
    \mathbf{h}_{u,m}[n] \sim \mathcal{CN}(\mathbf{0},\mathbf{R}_{u,m}),
    \label{eq:1} 
\end{equation}
where $\mathbf{R}_{u,m} \in \mathbb{C}^{L \times L}$ refers to the spatial correlation matrix of the Rayleigh-distributed non-line-of-sight (NLoS) components. The corresponding large-scale fading (LSF) coefficient, encompassing path loss, is denoted by $\alpha_{u,m} = \textrm{tr}(\mathbf{R}_{u,m})/L$. 

For the device-AP link, we also adopt a Rayleigh modeling:
\begin{equation}
    \mathbf{g}_{d,m}[n] \sim \mathcal{CN}(\mathbf{0},\mathbf{Q}_{d,m}),
    \label{eq:2} 
\end{equation}
with $\mathbf{Q}_{d,m} \in \mathbb{C}^{L \times L}$ the correlation in the NLoS propagation and $\beta_{d,m} = \textrm{tr}(\mathbf{Q}_{d,m})/L$ the set of LSF parameters.

\subsection{Uplink Channel Estimation}\label{sec:2.2}
Assuming perfect channel state information (CSI) can be overly optimistic in many applications. In practice, it is more realistic to obtain this information locally at the APs through UL orthogonal pilots\footnote{\color{black} In real-world scenarios, the low activities from mMTC+ allow pilot-based estimations (the number of active devices is comparable to the eMBB+ users).}. This approach enables the characterization of the sufficient statistics of the channels \cite{Ngo17}.

A feasible option could be the minimum mean-square error (MMSE) estimation \cite[Subsection~V-B]{Lie24}:
\begin{equation}
    \hat{\mathbf{h}}_{u,m}[n] = \sqrt{\eta_u} \mathbf{A}_{u,m} \bm{\varphi}_{u,m}[n],
    \label{eq:3}
\end{equation}
where $\eta_u$ ($\zeta_d$) is the training power, $\mathbf{A}_{u,m} \triangleq \mathbf{R}_{u,m} \mathbf{C}_{u,m}^{-1}$ refers to the MMSE matrix, and
\begin{equation}    
    \mathbf{C}_{u,m} = \sum_{k = 1}^{K_u} \eta_k \mathbf{R}_{k,m} \left\vert \bm{\phi}_k^{\textrm{H}} \bm{\phi}_u \right\vert^2 + \sum_{d = 1}^{K_d} \zeta_d \mathbf{Q}_{d,m} \left\vert \bm{\pi}_d^{\textrm{H}} \bm{\phi}_u \right\vert^2 + \sigma_m^2 \mathbf{I}_L,        
    \label{eq:4}
\end{equation}
denotes the covariance matrix of the least-squares observations $\bm{\varphi}_{u,m} [n] \in \mathbb{C}^L$ providing sufficient statistics, i.e.,
\begin{equation}
    \begin{aligned}
        \bm{\varphi}_{u,m}[n] &= \sqrt{\eta_u} \mathbf{h}_{u,m}[n] + \sum\nolimits_{k \neq u} \sqrt{\eta_k} \mathbf{h}_{k,m}[n] \bm{\phi}_k^{\textrm{H}} \bm{\phi}_u \\
        &\quad +  \sum\nolimits_{d = 1}^{K_d} \sqrt{\zeta_d} \mathbf{g}_{d,m}[n] \bm{\pi}_d^{\textrm{H}} \bm{\phi}_u + \bm{\omega}_{u,m} [n],
    \end{aligned}
    \label{eq:5}
\end{equation}
with $\bm{\phi}_u \in \mathbb{C}^{\tau_p}$ ($\bm{\pi}_d \in \mathbb{C}^{\tau_p}$) the sequence with length $\tau_p$ pilots sent by user $u$ (device $d$) to estimate the channel in all PRBs; and $\bm{\omega}_{u,m} [n] \in \mathbb{C}^L$ the ambient noise with variance $\sigma_m^2$. Please see \cite[Subsection~II-B]{Int23} for more details. 

The expressions for the mMTC+ estimates can be obtained similarly, yet are not included to avoid redundancy.

\subsection{Uplink Data Transmission} \label{sec:2.3}
Accordingly, the signal received at the $m$-th AP reads as
\begin{equation}    
    \mathbf{r}_m[n] =  \mathbf{y}_m^{\textrm{eMBB+}}[n] + \mathbf{y}_m^{\textrm{mMTC+}}[n] + \mathbf{w}_m[n],        
    \label{eq:6}
\end{equation}
where $\mathbf{w}_m[n]$ represents the additive white Gaussian noise, i.e., $\mathbf{w}_m[n] \sim \mathcal{CN}(\mathbf{0}_L,\sigma_m^2 \mathbf{I}_L)$.

The first term in \eqref{eq:6} contains the information of the eMBB+ terminals and can be modeled as follows:
\begin{equation}    
    \mathbf{y}_m^{\textrm{eMBB+}} [n] = \sum\nolimits_{u = 1}^{K_u} \sqrt{p_u}\mathbf{h}_{u,m}[n]s_u[n]            
    \label{eq:7}
\end{equation}
where $p_u$ is the UL power budget, $\mathbf{h}_{u,m}[n] \in \mathbb{C}^L$ is the user-AP channel from \eqref{eq:1}, and $s_u[n]$ is the transmit signal of user $u$. We also consider that all $s_u[n]$ are normally distributed with zero mean and unit power, i.e., $s_u[n] \sim \mathcal{CN}(0,1)$.

The second term comprises the mMTC+ messages, i.e.,
\begin{equation}
    \mathbf{y}_m^{\textrm{mMTC+}} [n] = \sum\nolimits_{d = 1}^{K_d} \sqrt{q_d} \mathbf{g}_{d,m}[n] x_d[n],
    \label{eq:8}
\end{equation}
where $q_d$ is the power coefficient and $\mathbf{g}_{d,m}[n]$ is the device-AP channel. In that sense, $x_d[n]$ is the signal transmitted by device $d$ and can be expressed as
\begin{equation}
    x_d[n] = c_d[n] z_d,
    \label{eq:9}
\end{equation}
where $c_d[n] \in \mathbb{C}$ denotes the unit-energy spreading waveform generated from a \textit{pseudo-noise} (PN) sequence comprising $T$ chips. As per the existing literature (cf. \cite{Mol22}), these sequences have cross-correlation factors of $1/T$. In this work, we further assume that each of them not only spans across time or frequency but across both resources simultaneously, i.e., $T = N$ (all the available PRBs). This implies that the data symbol $z_d$ will be common for all time-frequency slots. Like before, this signal follows a complex Gaussian distribution with zero mean and unit power, i.e., $z_d \sim \mathcal{CN}(0,1)$.

\subsubsection{Spatial Detection}
Using linear filters $\mathbf{f}_{u,m}[n] \in \mathbb{C}^L$ and $\mathbf{t}_{d,m}[n] \in \mathbb{C}^L$, each AP will detect its associated terminals and forward them to the CPU for retrieving signals $s_u[n]$ and $x_d[n]$. As mentioned earlier, in practical (scalable) CF deployments, only a subset of serving APs is associated with each user and device. To ease of notation, this will be indicated by the binary coefficients $a_{u,m}$ and $b_{d,m}$, activating whenever the terminal and AP are connected.

Based on that, the eMBB+ information is directly obtained via aggregation, i.e., $\hat{s}_u[n] = \sum_{m=1}^M a_{u,m} \mathbf{f}_{u,m}^{\textrm{H}}[n] \mathbf{r}_m[n]$, one for each time-frequency resource. In contrast, due to the use of PN sequences for transmitting mMTC+ signals, we still need an extra step to recover the original symbols $z_d$. 

\subsubsection{Time-Frequency Despreading}
Once the received signal is equalized utilizing filters $\mathbf{t}_{d,m}[n]$, each AP will stack all its replicas $\tilde{x}_{d,m}[n] = \mathbf{t}_{d,m}^{\textrm{H}}[n] \mathbf{r}_m[n]$ into a large column vector $\tilde{\mathbf{x}}_{d,m} = [\tilde{x}_{d,m}[1],\ldots,\tilde{x}_{d,m}[N]]^{\textrm{T}}$ for later correlating it with the (modified) signature PN sequences, i.e.,
\begin{equation}
    \tilde{z}_{d,m} = \mathbf{c}_d^{\textrm{H}} \textrm{diag}\left(\bm{\mu}_{d,m}\right) \tilde{\mathbf{x}}_{d,m},
    \label{eq:10}
\end{equation}
where $\mathbf{c}_d = [c_d[1],\ldots,c_d[N]]^{\textrm{T}}$ is the original spreading waveform and the vector $\bm{\mu}_{d,m} = [\mu_{d,m}[1],\ldots,\mu_{d,m}[N]]^{\textrm{T}}$ concatenates the set of processed gains
\begin{equation}
    \mu_{d,m}[n] = \mathbf{g}_{d,m}^{\textrm{H}}[n] \mathbf{t}_{d,m}[n].
    \label{eq:11}
\end{equation}

To some extent, \eqref{eq:10} can be interpreted as a \textit{time-frequency despreading} \cite{Sim94}. Nevertheless, unlike single-antenna systems, we must introduce the weights $\mu_{d,m}[n]$ in \eqref{eq:10} to incorporate the benefits of spatial diversity. Otherwise, the new \textit{degrees of freedom} can hinder the spreading gain: (auto-/cross-) correlation properties of PN sequences would be lost.

Note that, for further suppressing mMTC+ interference, one can design the operation \eqref{eq:10} in a \mbox{zero-forcing (ZF) fashion}. \mbox{In a} nutshell, rather than simply employing $\bm{c}_d$ (which indeed resembles a matched filter), the correlating sequence of device $d$ can be projected onto the null space spanning the rest of the signatures \cite{Sim94}. Such analysis is beyond the scope of this investigation and will be conducted in future research lines.

Consequently, regardless of the strategy, the CPU will end up with a series of effective estimates $\hat{z}_d = \sum_{m=1}^M b_{d,m} \tilde{z}_{d,m}$ that are used to decode the mMTC+ data.

\section{Achievable Data Rate} \label{sec:3}
In the upcoming section, we formulate the power control to maximize the minimum data rate of the eMBB+ users subject to QoS constraints on the mMTC+ devices. To do so, we will first report an achievable lower bound on the data rates with channel estimation errors for both terminals.

In the presence of imperfect CSI, one can obtain a tractable expression for the data rate through the \textit{use-and-then-forget} (UatF) bound, widely used in mMIMO. Essentially, it implies that channel estimates are only exploited for beamforming and later dumped during signal detection (cf. \cite[(11)]{Int23}).

Concisely, assuming the CPU only has channel distribution information (CDI) and no knowledge of the realizations, the received signal of user $u$ can be expressed as in \eqref{eq:12} at the top of the next page, where the estimation errors are treated as an additional ``effective" noise $U_u[n]$. Following the rationale in \cite[Theorem 5.4]{Dem21}, we will consider the worst-case scenario and model $U_u[n]$ as uncorrelated Gaussian noise. This will lead to a lower bound on the (ergodic) channel capacity.

\begin{figure*}[t]
\begin{align}     
    \hat{s}_u [n] &= \sqrt{p_u} \underbrace{\mathbb{E}\left[\sum_m a_{u,m} \mathbf{f}_{u,m}^{\textrm{H}}[n] \mathbf{h}_{u,m}[n] \right]}_{\triangleq D_u}s_u[n] + \sqrt{p_u} \underbrace{\left(\sum_m a_{u,m} \mathbf{f}_{u,m}^{\textrm{H}}[n] \mathbf{h}_{u,m}[n] - D_u\right)}_{\triangleq U_u[n]} s_u[n] + \underbrace{\sum_m a_{u,m} \mathbf{f}_{u,m}^{\textrm{H}}[n]\mathbf{w}_m[n]}_{\triangleq W_u^{\textrm{eMBB+}}[n]} \nonumber \\
    &\quad + \sum_{k \neq u} \sqrt{p_k} \underbrace{\sum\nolimits_m a_{u,m} \mathbf{f}_{u,m}^{\textrm{H}}[n] \mathbf{h}_{k,m}[n]}_{\triangleq I_{u,k}^{\textrm{eMBB+}}[n]} s_k[n]  + \sum_d \sqrt{q_d}  \underbrace{\sum\nolimits_m a_{u,m} \mathbf{f}_{u,m}^{\textrm{H}}[n] \mathbf{g}_{d,m}[n]}_{\triangleq I_{u,d}^{\textrm{mMTC+}}[n]} x_d[n],    
    \label{eq:12}
\end{align}
\vspace{-2mm}
\hrule
\end{figure*}

This way, the achievable data rate results in 
\begin{equation}
    R_u^{\textrm{eMBB+}}\left(\mathcal{P}\right) = \frac{\tau_u}{\tau_c} B \log_2\left(1 + \gamma_u\left(\mathcal{P}\right)\right),
    \label{eq:13}
\end{equation}
with $\tau_u$, $\tau_c$, and $B$ the number of UL transmit symbols, the number of (time-frequency) coherence samples per PRB, and the system's bandwidth, respectively. For clarity, $\mathcal{P}$ is the set containing the coefficients $p_u$ and $q_d$. Accordingly, the signal-to-interference-plus-noise ratio (SINR) is
\begin{equation}
    \gamma_u\left(\mathcal{P}\right) = \frac{\displaystyle p_u \delta_u}{\displaystyle p_u \upsilon_u + \sum\nolimits_{k \neq u}p_k \kappa_{u,k} + \sum\nolimits_d q_d \varkappa_{u,d} + \xi_u},
    \label{eq:14}
\end{equation}
where $\delta_u = \left\vert D_u \right\vert^2$ is the strength of the useful signal and $\upsilon_u = \mathbb{E}[\vert U_u[n] \vert^2]$ represents the channel uncertainty, whilst $\kappa_{u,k} = \mathbb{E}[\vert I_{u,k}^{\textrm{eMBB+}} [n] \vert^2]$, $\varkappa_{u,d} = \mathbb{E}[\vert I_{u,d}^{\textrm{mMTC+}} [n]\vert^2]$, and $\xi_u = \mathbb{E}[\vert W_u^{\textrm{eMBB+}}[n] \vert^2 ]$ refer to the powers of the (eMBB+/mMTC+) interference and noise, respectively. As we will see, the key idea here is that the mMTC+ contribution is almost negligible, since the transmit powers $q_d$ of the devices will be very small compared to those coefficients $p_u$ of the eMBB+ users. This is experimentally verified in Section~\ref{sec:5}.

Similarly, to derive the UatF bound for device $d$, we will first express the received signal as in \eqref{eq:15} given at the top of page 5, where $\tilde{\mathbf{t}}_{d,m}[n] = \mathbf{g}_{d,m}^{\textrm{H}}[n] \mathbf{t}_{d,m}[n] \mathbf{t}_{d,m}^{\textrm{H}}[n]$. Once again, we introduce the impact of the lack of CSI via an effective noise $V_u$. Note that, different from \eqref{eq:12}, where each user experiences $N$ different realizations over the resource grid, expression \eqref{eq:15} is unique for each device (cf. \eqref{eq:10}). 

As a result, we can show that the corresponding SINR yields%
\setcounter{equation}{15}
\begin{equation}
    \rho_d\left(\mathcal{P}\right) = \frac{\displaystyle q_d \lambda_d}{\displaystyle q_d \nu_d + \sum\nolimits_{k \neq d}q_k \epsilon_{d,k} + \sum\nolimits_u p_u \varepsilon_{d,u} + \chi_d},
    \label{eq:16}
\end{equation}
where $\lambda_d = \vert S_d \vert^2$ denotes the power of the desired signal, $\nu_d \mathbb{E}[\vert V_d \vert^2]$ comprises the effect of imperfect CSI, and $\epsilon_{d,k} = \mathbb{E}[\vert J_{d,k}^{\textrm{mMTC+}} \vert^2]$, $\varepsilon_{d,u} = \mathbb{E}[\vert J_{d,u}^{\textrm{eMBB+}} [n]\vert^2]$, $\chi_d = \mathbb{E}[\vert W_d^{\textrm{mMTC+}} \vert^2]$ are the strength of the (mMTC+/eMBB+) interfering signals and thermal noise, respectively. Unlike \eqref{eq:14}, now the focus lies on the so-called spreading gain $N$ for the mMTC+ devices. A more comprehensive discussion is provided in Section~\ref{sec:5}.

The closed forms of the moments above are difficult to find. However, under maximum ratio combining, i.e., $\mathbf{f}_{u,m}[n] = \hat{\mathbf{h}}_{u,m}[n]$ ($\mathbf{t}_{u,m}[n] = \hat{\mathbf{g}}_{u,m}[n]$), they can be obtained after some manipulations. In the eMBB+ case, we have \cite[Corollary~2]{Bjo20}
\begin{equation}
    \begin{aligned}
        \delta_u &= \eta_u^2 \left\vert \sum\nolimits_m a_{u,m} \textrm{tr}\left(\mathbf{A}_{u,m} \mathbf{R}_{u,m}\right) \right\vert^2, \\        
        \upsilon_u &= \eta_u \sum\nolimits_m a_{u,m} \textrm{tr}\left(\mathbf{A}_{u,m} \mathbf{R}_{u,m} \mathbf{R}_{u,m}\right), \\
        \kappa_{u,k} &= \eta_u \sum\nolimits_m a_{u,m} \textrm{tr}\left(\mathbf{A}_{u,m} \mathbf{R}_{u,m} \mathbf{R}_{k,m}\right) \\
        &\quad + \eta_u \eta_k \left\vert \bm{\phi}_k^{\textrm{H}} \bm{\phi}_u \right\vert^2 \left\vert \sum\nolimits_m a_{u,m} \textrm{tr}\left(\mathbf{A}_{u,m} \mathbf{R}_{k,m}\right) \right\vert^2, \\
        \varkappa_{u,d} &= \eta_u \sum\nolimits_m a_{u,m} \textrm{tr}\left(\mathbf{A}_{u,m} \mathbf{R}_{u,m} \mathbf{Q}_{d,m}\right) \\
        &\quad + \eta_u \zeta_d \left\vert \bm{\pi}_d^{\textrm{H}} \bm{\phi}_u \right\vert^2 \left\vert \sum\nolimits_m a_{u,m} \textrm{tr}\left(\mathbf{A}_{u,m} \mathbf{Q}_{d,m}\right) \right\vert^2, \\
        \xi_u &= \eta_u \sum\nolimits_m a_{u,m} \sigma_m^2 \textrm{tr}\left(\mathbf{A}_{u,m} \mathbf{R}_{u,m}\right),
    \end{aligned}    
    \label{eq:17}
\end{equation}
whereas for the mMTC+ devices, the computations involving high-order moments might become tedious, e.g., expectations of 4 inner/outer products ($8$-th order) \cite{Mal11}. These can be easily circumvented using numerical evaluation (i.e., approximating statistical averages by sample means \cite{Ozd19}), yet when particularizing for uncorrelated fading, the following can be reported:
\begin{equation}
        \lambda_d = N^2L^2\left\vert \sum\nolimits_m b_{d,m} \hat{\beta}_{d,m}^2 \tilde{\beta}_{d,d,m} \left(\bar{\beta}_{d,m} + L \tilde{\beta}_{d,d,m} \right) \right\vert^2,            
    \label{eq:18}
\end{equation}
with $\hat{\beta}_{d,m} = \textrm{tr}(\mathbf{B}_{d,m})/L$ for convenience and $\mathbf{B}_{d,m} \in \mathbb{C}^{L \times L}$ the MMSE matrix for device $d$ (cf. \eqref{eq:3}). Analogously, we also define $\bar{\beta}_{d,m} = \sum_k \tilde{\beta}_{k,d,m} + \sum_u \tilde{\alpha}_{u,d,m} + \sigma_m^2$, where $\tilde{\beta}_{k,d,m} = \zeta_k \beta_{k,m} \vert \bm{\pi}_k^{\textrm{H}} \bm{\pi}_d\vert^2$ and $\tilde{\alpha}_{u,d,m} = \eta_u \alpha_{u,m} \vert \bm{\phi}_u^{\textrm{H}} \bm{\pi}_d \vert^2$. The other terms in \eqref{eq:16} are given in \eqref{eq:19} at the top of page 6. 

{\color{black} Note that, while \eqref{eq:17} extends the typical eMBB+ framework to incorporate mMTC+ interference, at the time of writing, \eqref{eq:18} and \eqref{eq:19} are novel in CF-mMIMO literature. This opens the door to breakthroughs for coexisting 6G technologies.} 

To meet the page limit, the detailed procedure is omitted. In addition, the derivations including more advanced processing techniques such as ZF and MMSE are left for future studies. Interested readers might refer to \cite{Bjo20} and references therein.

\begin{figure*}[t]
\begin{align}    
    \hat{z}_d &= \sqrt{q_d} \underbrace{\mathbb{E}\left[\sum_{n,m} b_{d,m} \left\vert \mathbf{t}_{d,m}^{\textrm{H}}[n] \mathbf{g}_{d,m}[n] \right\vert^2 \right]}_{\triangleq S_d}z_d + \sqrt{q_d} \underbrace{\left( \sum_{n,m} b_{d,m} \left\vert \mathbf{t}_{d,m}^{\textrm{H}}[n] \mathbf{g}_{d,m}[n] \right\vert^2 - S_d\right)}_{\triangleq V_d} z_d + \underbrace{\sum_{n,m} c_d^*[n] b_{d,m} \tilde{\mathbf{t}}_{d,m}^{\textrm{H}}[n] \mathbf{w}_m[n]}_{\triangleq W_d^{\textrm{mMTC+}}} \nonumber \\      
    &\quad + \sum_{k \neq d} \sqrt{q_k} \underbrace{\sum_n c_d^*[n] c_k[n] \sum_m b_{d,m} \tilde{\mathbf{t}}_{d,m}^{\textrm{H}}[n] \mathbf{g}_{k,m}[n]}_{\triangleq J_{d,k}^{\textrm{mMTC+}}} z_k  + \sum_u \sqrt{p_u} \sum_n \underbrace{c_d^*[n] \sum_m b_{d,m} \tilde{\mathbf{t}}_{d,m}^{\textrm{H}}[n] \mathbf{h}_{u,m}[n]}_{\triangleq J_{d,u}^{\textrm{eMBB+}} [n]} s_u[n],     
    \label{eq:15} \tag{15}
\end{align}
\vspace{-2mm}
\hrule
\vspace{-4mm}
\end{figure*}

\section{Power Control Design} \label{sec:4}
At this point, we can formulate the following optimization, in which $C1$ and $C2$ constrain the transmit powers, $C3$ ensures all devices have rates above a certain QoS threshold, whereas $C4$ guarantees a minimum SINR for a reliable mMTC+ packet decoding (otherwise, too many decoding errors might occur, leading to possible communication failure):%
\setcounter{equation}{19}
\begin{equation}
    \begin{aligned}
        \underset{\mathcal{P}}{\textrm{max}} \, \underset{u}{\textrm{min}} \: & R_u^{\textrm{eMBB+}} \left(\mathcal{P}\right) & \\
        \textrm{s.t.} \quad &C1: 0 \leq p_u \leq P_u \; \forall u &C2: 0 \leq q_d \leq P_d  \; \forall d\\
        &C3:  R_d^{\textrm{mMTC+}} \left(\mathcal{P}\right) \geq r \; \forall d &C4:  \rho_d \left(\mathcal{P}\right) \geq s  \; \forall d, \: \, 
    \end{aligned}
    \label{eq:20}
\end{equation}
with $R_d^{\textrm{mMTC+}}(\mathcal{P}) = \frac{\tau_u}{\tau_c} \frac{B}{N} \log_2(1 + \rho_d(\mathcal{P}))$, where $1/N$ is the cost of the proposed spreading, i.e., using the PN sequences requires $N$-fold more (time/frequency) samples. {\color{black} Thus, unlike SINRs, the ``penalized'' mMTC+ rates will decrease with $N$.}

Equivalently, the problem defined in \eqref{eq:20} can be translated into the following standard epigraph form \cite[(4.11)]{Boy04}:
\begin{equation}
    \begin{aligned}
        \underset{\mathcal{P},t}{\textrm{max}} \quad t \quad \textrm{s.t.} \quad & C1, \ldots, C4 \\
        &C5:  R_u^{\textrm{eMBB+}}\left(\mathcal{P}\right) \geq t \quad \forall u,
    \end{aligned}
    \label{eq:21}    
\end{equation}
which can be cast as a quasi-linear problem \cite{Ngo17}. The proof is two-fold: (i) $C1$, $C2$, and $C4$ are linear by definition; (ii) $C3$ and $C5$ can also be written as linear constraints thanks to the monotonically increasing nature of the logarithm. 

As a result, the global optimum of problem \eqref{eq:21} can feasibly be found by using a bisection search and solving a sequence of linear feasibility problems \cite[Algorithm~4.1]{Boy04}. {\color{black} In line with the CF goal, this power control method will contribute to terminals having uniformly good QoS over the entire service area.}

\begin{figure}[t]
    \centerline{\includegraphics[scale = 0.975]{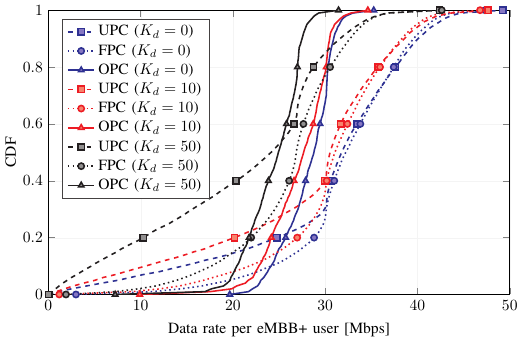}}  
    \vspace{-1mm}
    \caption{Rate of eMBB+ users vs. number of mMTC+ devices $K_d$.}
    \label{fig:2}
    \vspace{1mm}
    \centerline{\includegraphics[scale = 0.975]{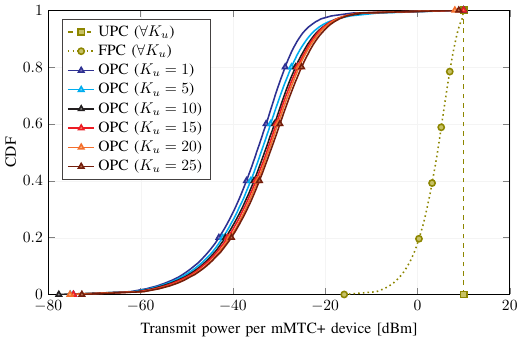}}
    \vspace{-1mm}
    \caption{Power of mMTC+ devices vs. number of eMBB+ users $K_u$.}
    \label{fig:3}
    \vspace{-5mm}
\end{figure}

\begin{figure}[t]
    \centerline{\includegraphics[scale = 0.975]{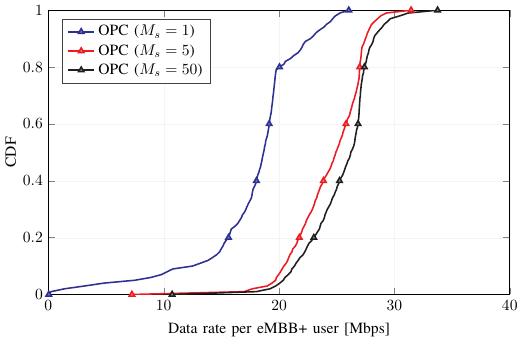}}    
    \vspace{-1mm}
    \caption{Rate of eMBB+ users vs. number of serving APs $M_s$.}
    \label{fig:4}
    \vspace{1mm}
    \centerline{\includegraphics[scale = 0.975]{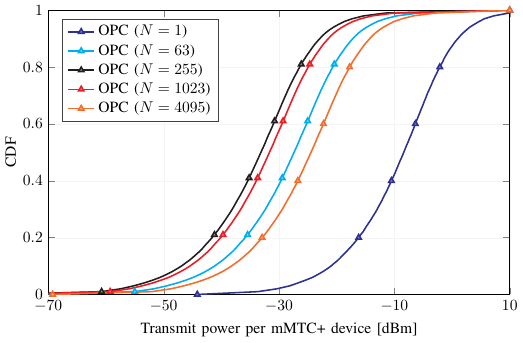}}
    \vspace{-1mm}
    \caption{Power of mMTC+ devices vs. number of PRBs $N$.}
    \label{fig:5}
    \vspace{-5mm}
\end{figure}

\begin{figure*}[t]
\begin{align}
    \nu_d &=  N L(L + 1) \sum\nolimits_m b_{d,m} \hat{\beta}_{d,m}^4 \tilde{\beta}_{d,d,m}^2 \left(L\left(L + 1\right) \tilde{\beta}_{d,d,m}^2 + 2 \bar{\beta}_{d,m} \left(2\left(L + 1\right)\tilde{\beta}_{d,d,m} + \bar{\beta}_{d,m}\right) - L\left(\bar{\beta}_{d,m} + L \hat{\beta}_{d,m}\right)^2\right), \nonumber \\
    \epsilon_{d,k} &=  N L \zeta_d \sum\nolimits_m b_{d,m} \hat{\beta}_{d,m}^4 \tilde{\beta}_{d,d,m} \beta_{k,m} \left(\left(L + 1\right)\bar{\beta}_{d,m}^2 + \left(L + 1\right)^2\bar{\beta}_{d,m}\left(\tilde{\beta}_{d,d,m} + \tilde{\beta}_{k,d,m}\right) + L\left(2L + 1\right) \tilde{\beta}_{d,d,m} \tilde{\beta}_{k,d,m}\right) \nonumber \\
    &\quad + L^4\zeta_d \zeta_k \left\vert \bm{\pi}_k^{\textrm{H}} \bm{\pi}_d \right\vert^2 \left\vert \sum\nolimits_m b_{d,m} \hat{\beta}_{d,m}^2 \tilde{\beta}_{d,d,m} \beta_{k,m} \right\vert^2, \nonumber \\
    \varepsilon_{d,u} &= NL \zeta_d \sum\nolimits_m b_{d,m} \hat{\beta}_{d,m}^4 \tilde{\beta}_{d,d,m} \alpha_{u,m} \left(\left(L + 1\right)\bar{\beta}_{d,m}^2 + \left(L + 1\right)^2\bar{\beta}_{d,m}\left(\tilde{\beta}_{d,d,m} + \tilde{\alpha}_{u,d,m}\right) + L\left(2L + 1\right) \tilde{\beta}_{d,d,m} \tilde{\alpha}_{u,d,m}\right) \nonumber \\
    &\quad + NL^4\zeta_d \eta_u \left\vert \bm{\phi}_u^{\textrm{H}} \bm{\pi}_d \right\vert^2 \left\vert \sum\nolimits_m b_{d,m} \hat{\beta}_{d,m}^2 \tilde{\beta}_{d,d,m} \alpha_{u,m} \right\vert^2, \nonumber \\
    \chi_d &= NL(L + 1) \zeta_d \sum\nolimits_m b_{d,m}  \hat{\beta}_{d,m}^4 \tilde{\beta}_{d,d,m} \bar{\beta}_{d,m} \sigma_m^2\left(\left(L + 1\right)\tilde{\beta}_{d,d,m} + \bar{\beta}_{d,m}\right),
    \label{eq:19} \tag{19} 
\end{align}
\vspace{-2mm}
\hrule
\vspace{-4mm}
\end{figure*}

\section{Numerical Simulations} \label{sec:5}
In what follows, we present numerous experiments to validate the results and assess the performance of our approach. Namely, we plot the cumulative distribution functions (CDFs) of the eMBB+ data rates and the mMTC+ transmit powers.

Across all simulations, users, devices, and APs will be uniformly located inside a deployment area of $1$ km\textsuperscript{2}, wrapped around the edges to avoid possible boundary effects. Regarding the terminal-AP association, we assume each is linked to the $M_s = 5$ APs with the largest LSF coefficients \cite{Elw23}.

The scenario follows the micro-urban configuration from \cite{3GPP36814} with $P_u = 20$ dBm $\ll$ $P_d = 10$ dBm $\forall u, d$, $\sigma_m^2 = N_o B$ $\forall m$, $N_o = -174$ dBm/Hz, and $B = 20$ MHz. The carrier frequency is $2$ GHz and the set of PN signatures are generated according to \textit{m-sequences} with $N = 2^n - 1$ for $n = 1,2,\ldots$ \cite[Section~5.4]{Sim94}. We also consider PRBs of size $1$ ms and $200$ kHz \cite{Elw23}, i.e., $\tau_c = 200$ resource elements. The first $\tau_p = (K_u + K_d)/2$ symbols are dedicated to channel estimation, and $\tau_u = (\tau_c - \tau_p)/2$ are assigned to UL communication. 

Unless otherwise stated, we fix $K_u = 10 \ll K_d = 50$, $L = 8$, $M = 50$, $N = 255$, and $r = 10$ kbps\footnote{mMTC+ are characterized by sporadic transmissions (the number of active devices $K_d$ is much less than the total number $D$). For instance, for periodic reports of $l = 10$ kb every $t = 6$ h, $K_d = 50$ is equivalent to supporting $D = (rt/l) K_d \approx 10^6 $ terminals over the $1$ km\textsuperscript{2} deployment area.}. Lastly, recall that 3GPP suggests low-order constellations for mMTC+. QPSK modulations need SINRs of $s = -6.7$ dB to achieve block error probabilities less than $10\%$ \cite[Table~4.7]{Gho11}.

Together with the optimal power control (OPC) derived in Section~\ref{sec:4}, we will include uniform power control (UPC) and fractional power control (FPC) as benchmark schemes \cite{Dem21}. We modify both techniques so that they also comply with all constraints $C1,\ldots,C4$ for a fair comparison.

In Figs.~\ref{fig:2} and \ref{fig:3}, we plot the eMBB+ data rate (UL) and the mMTC+ transmit power with respect to (w.r.t.) the number of devices $K_d$ and users $K_u$, respectively. As expected, when introducing mMTC+ devices in the scenario, the performance of eMBB+ users degrades (more interference is experienced). This clearly means the coexistence of services poses a limit on the attainable QoS since denser mMTC+ deployments might prevent feasible eMBB+ communication. Additionally, OPC yields a better minimum user rate in all cases, especially when $K_d$ increases. In short, the system becomes more sensitive to interference and the optimal solution becomes more restrictive.

On the other hand, higher transmit powers are needed to cope with increasing user traffic loads. This is not surprising as mMTC+ face more eMBB+ interference and adjust their coefficients accordingly. Remarkably, though, the difference is quite small (in the case of UPC and FPC, little to none). This highlights the robustness of our design against $K_u$, i.e., we can handle more eMBB+ users almost effortlessly. As before, OPC outperforms the other strategies (far less power is used). To avoid redundancy, we now concentrate on OPC.

The eMBB+ performance w.r.t. the number of serving APs is depicted in Fig.~\ref{fig:4}. Note that, $M_s = 1$ resembles a small-cell operation \cite{Ngo17} while $M_s = 5$ corresponds to our TC-CF deployment and $M_s = M$ equals a pure CF network. One can see that larger values of $M_s$ yield better data rates, which justifies the use of CF over cellular systems. However, by increasing $M_s$, the required fronthaul and computational complexity also grow unboundedly (making the design unfeasible in practice). That is why in this work, to achieve scalability, we advocate for a TC implementation. The gap is significant when compared to the small-cell approach, but improvement rapidly saturates for higher numbers of serving APs.

Finally, Fig.~\ref{fig:5} illustrates the device power w.r.t. the number of PRBs $N$. We observe that a low $N$ entails higher powers, as the SINR would be otherwise too low for correct decoding ($C4$). The case of no spreading, i.e., $N = 1$, indeed requires the largest transmit power. For high numbers of PRBs, however, large values of power are also needed to compensate for the effect of the penalty coefficient $1/N$ in the pre-log term of the mMTC+ devices rate. The figure thus reveals that there is an optimal value for $N$ that minimizes the power required for the mMTC+ devices to achieve their desired target rate ($C3$). In particular, the figure shows that for the case at hand, the value $N=255$ achieves the lowest values of transmit power. 

\section{Conclusions} \label{sec:6}
This paper has addressed the problem of designing proper MA strategies for coexistence between eMBB+ and mMTC+ data services in a (novel) terminal-centric CF-mMIMO network. The use of a time-frequency spreading technique for the mMTC+ devices has been proposed, along with suitable QoS-based power control mechanisms, to efficiently multiplex the two types of traffic in the same resource grid of time-frequency PRBs. Simulations have shown that the proposed multiplexing scheme is effective and attains satisfactory performance levels.

\bibliographystyle{IEEEtran}
\bibliography{references}

\end{document}